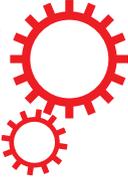

# OPEN
# The sequence to hydrogenate coronene cations: A journey guided by magic numbers

Stéphanie Cazaux[1,2], Leon Boschman[1,3], Nathalie Rougeau[4], Geert Reitsma[3,5], Ronnie Hoekstra[3], Dominique Teillet-Billy[4], Sabine Morisset[4], Marco Spaans[1] & Thomas Schlathölter[3]

The understanding of hydrogen attachment to carbonaceous surfaces is essential to a wide variety of research fields and technologies such as hydrogen storage for transportation, precise localization of hydrogen in electronic devices and the formation of cosmic $H_2$. For coronene cations as prototypical Polycyclic Aromatic Hydrocarbon (PAH) molecules, the existence of magic numbers upon hydrogenation was uncovered experimentally. Quantum chemistry calculations show that hydrogenation follows a site-specific sequence leading to the appearance of cations having 5, 11, or 17 hydrogen atoms attached, exactly the magic numbers found in the experiments. For these closed-shell cations, further hydrogenation requires appreciable structural changes associated with a high transition barrier. Controlling specific hydrogenation pathways would provide the possibility to tune the location of hydrogen attachment and the stability of the system. The sequence to hydrogenate PAHs, leading to PAHs with magic numbers of H atoms attached, provides clues to understand that carbon in space is mostly aromatic and partially aliphatic in PAHs. PAH hydrogenation is fundamental to assess the contribution of PAHs to the formation of cosmic $H_2$.

Molecular hydrogen is the most abundant molecule of our Universe. Next to dust, polycyclic aromatic hydrocarbons (PAHs) can be important catalysts to the formation of $H_2$[1]. The presence of pre-adsorbed H atoms on PAHs increases the yield of $H_2$ formation by many orders of magnitude[2], rendering this process an important route towards molecular hydrogen formation in the interstellar medium (ISM)[3].

A multitude of objects inside and outside our galaxy exhibit spectra crowded with infrared bands, referred to as the Aromatic Infrared Bands (AIBs). These AIBs are commonly attributed to PAH molecules[4,5]. To date, families of these astronomical PAHs have been identified[6,7], in neutral or cationic forms[8], being predominantly aromatic[9] and likely hydrogenated[10]. However, a unique identification is hampered by the non specific nature of the IR bands (depicting vibrations of functional groups) and the large number of candidate PAH molecules.

Hydrogenated graphene-related materials are also considered as a successor of silicon in the electronics industry. However, the zero band-gap of pure graphene prohibits the direct employment of its excellent conductive properties in semiconductor devices. Hydrogen adsorption on graphene changes the state of the functionalized C atom from $sp^2$ to $sp^3$ and removes electrons from the valence shells, opening up the graphene band-gap in a tunable fashion[11] which is a prerequisite for graphene applications in electronic circuits.

Despite the enormous interest in hydrogenation of graphene related materials, with graphene, graphite and PAHs being the most widely studied model systems, the key adsorption process of a single H atom on a carbonaceous surface is still not fully understood. For instance, the exact height of the energy barrier a single H atom encounters upon adsorbing on the surface is still under debate. Density functional theory (DFT) calculations predicted an H chemisorption barrier of 0.2 eV for graphite[12,13] implying that the hydrogenation process is very unlikely at low temperatures. Experiments confirmed that chemisorption barriers could be overcome by thermal H atoms ($\geq 2000$ K)[14], but showed that one already adsorbed hydrogen atom decreases the barrier for chemisorption of a second H atom[15,16] allowing for barrier-less $H_2$ formation and the formation of hydrogen clusters[17,18]. Also, recent studies indicate that both van der Waals and quantum nuclear effects cooperatively lower chemisorption barriers for hydrogenation on graphene and PAH molecules in a similar manner[19].

[1]Kapteyn Astronomical Institute, University of Groningen, P.O. Box 800, NL 9700 AV Groningen, The Netherlands. [2]Leiden Observatory, Leiden University, P.O. Box 9513, NL 2300 RA Leiden, The Netherlands. [3]Zernike Institute for Advanced Materials, University of Groningen, Nijenborgh 4, 9747AG Groningen, The Netherlands. [4]Institut des Sciences Moléculaires d'Orsay, ISMO, CNRS, Univ Paris-Sud, Université Paris Saclay, F-91405 Orsay, France. [5]Max-Born-Institute, Max Born Strasse 2A, D-12489 Berlin, Germany. Correspondence and requests for materials should be addressed to S.C. (email: cazaux@astro.rug.nl)

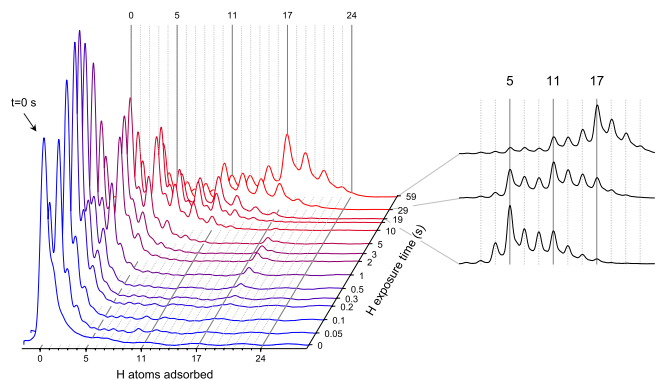

**Figure 1.** Panel **(a)** Mass spectrum of coronene after exposure with hydrogen atoms for different exposure times. Panel **(b)** mass spectra obtained obtained for $t_H$ = 19, 29 and 59 s, highlighting the presence of magic numbers.

Motivated by the fact that atomic hydrogen addition to graphene related materials strongly depends on the presence of already adsorbed H atoms, the electronic and structural effects of this process on the atomic level in an isolated system is investigated in this paper. Here experimental and theoretical studies are combined to study the sequential addition of thermal hydrogen atoms to an isolated, gas phase coronene cation as the prototypical PAH molecule. Briefly, we find that the hydrogenation of coronene cations follows definite pathways where barriers are highest for the hydrogenation of coronene cations with 5, 11 and 17 hydrogen atoms. These high barriers imply that coronene cations with magic numbers of H atoms attached (5, 11, and 17) are the most abundant. Quantum chemistry calculations confirm the experimental data and, moreover, show a site-specific sequence of hydrogenation. The establishment of the path to the most probable superhydrogenated PAH cations allows for predicting which PAH cations are more likely present in space and will advance the search for candidates of AIBs. The possibility of assembling gas phase PAH cations in well-defined hydrogenation states encourages the investigation of their electronic and geometric structures as model systems for graphene hydrogenation.

## Results

**Experimental results.** Gas phase coronene cations confined in a radiofrequency trap were exposed to a constant flux of H atoms for variable time intervals. By varying the irradiation time the degree of hydrogenation was changed. The data obtained from the experiment are a series of mass spectra of superhydrogenated coronene cations as a function of H exposure time. Fig. 1a) displays the evolution of the mass spectra for exposure times between 0 and 59 s, normalized to the total peak area. The reference spectrum with zero H atom exposure consists of a dominating peak at the nominal coronene m/z of 300, i.e. there is zero H atom adsorption ($n_H = 0$). The shoulder at m/z = 301 is due to the natural 26% contribution of $^{13}C$.

Upon H exposure, the *m/z* distribution exhibits the expected shift to higher masses, accompanied by a broadening of the distribution. Odd *m/z* values dominate the spectrum, with the peaks at even *m/z* being mainly caused again by the natural $^{13}C$ content of the sample. Note that the scale is logarithmic on both axes. It is obvious that brief H exposure for $t_H = 0.05$ s leads to single hydrogenation ($C_{24}H_{13}^+$, i.e. one additional H atom: $n_H = 1$) of more than 50% of the trap content. After $t_H = 0.2$ s, 90% of the trap content is at least singly hydrogenated ($n_H = 1$). The $C_{24}H_{13}^+$ ($n_H = 1$) peak becomes dominant for $t_H = 0.2 - 0.3$ s and then decreases by a factor of 20 after $t_H = 10$ s. With declining $C_{24}H_{13}^+$ ($n_H = 1$), $C_{24}H_{15}^+$ ($n_H = 3$) increases and becomes the dominant contribution at $t_H = 5$ s. From $t_H = 10$ s on, $C_{24}H_{17}^+$ ($n_H = 5$) becomes the dominant peak. This hydrogenation state is particular, as the two subsequent odd hydrogenation states ($C_{24}H_{19}^+$ and $C_{24}H_{21}^+$ which corresponds to $n_H = 7$ and 9 respectively) have markedly lower yields over the full range. A similar behaviour is observed from $t_H = 20$ s on, where $C_{24}H_{23}^+$ ($n_H = 11$) exceeds the two subsequent odd hydrogenation states ($C_{24}H_{25}^+$ and $C_{24}H_{27}^+$ which corresponds to $n_H = 13$ and 15 respectively). Eventually, the $C_{24}H_{29}^+$ ($n_H = 17$) peak becomes the most important one, once again exceeding the yields of neighboring odd hydrogenation states. For exposure times around 60 s, full hydrogenation ($n_H = 23$, odd hydrogenation state before full hydrogenation $n_H = 24$) starts to contribute to the mass spectra.

To illustrate the "magic" nature of the hydrogenation states $n_H = 5$, 11 and 17 even more clearly, Fig. 1b) shows the data obtained for $t_H = 19$, 29 and 59 s, separately. Clear steps in peak intensity occur at the magic hydrogenation states $n_H = 5$ and $n_H = 11$ in all three spectra, while $n_H = 17$ dominates for the longest exposure time.

These findings suggest particularly high stabilities of coronene cations with $n_H = 5$, 11 and 17 and particularly high barriers for the transitions to the respective subsequent hydrogenation states ($n_H = 5 \rightarrow 6$, $n_H = 11 \rightarrow 12$ and $n_H = 17 \rightarrow 18$).

**Theoretical results.** DFT calculations were performed to clarify the origin of the experimentally observed magic numbers. Binding energies for relevant superhydrogenated cations and barriers for various pathways of interest have been calculated with the modified MPW1K hybrid functional[20]. These calculations predict a well defined hydrogenation pathway from the singly hydrogenated cation $C_{24}H_{13}^+$ to the fully hydrogenated cation $C_{24}H_{36}^+$.

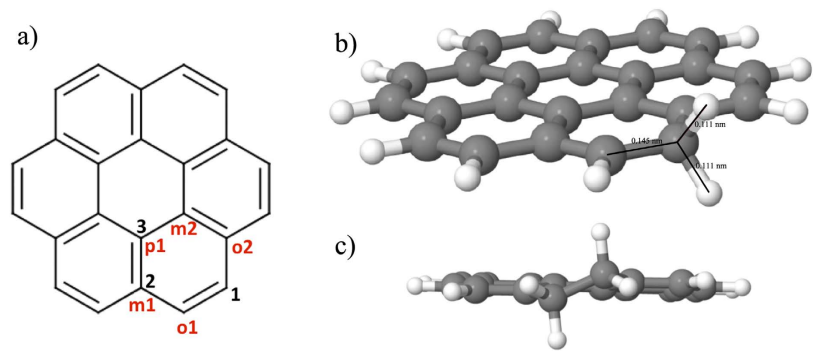

**Figure 2. Left panel: distinct type of hydrogenation sites considered in our calculations.** Sites for the first hydrogenation are represented in black while sites for the second hydrogenation (with first H located in 1) are represented in red. Right panel Top: Geometry of the coronene cation with an additional H atom adsorbed on an outer edge carbon. Right panel bottom: Geometry of the coronene cation with two additional H atoms adsorbed on adjacent outer edge carbons.

| | First hydrogenation | | | Second hydrogenation | | |
|---|---|---|---|---|---|---|
| Site | $\Delta E_b$ (ZPE corrected) (eV) Present work (MPW1K) | $\Delta E_b$ (eV) [ref. 21] B3LYB | Barrier (ZPE corrected) (eV) Present work (MPW1K) | Site | $\Delta E_b$ (eV) | Barrier (eV) |
| 1 | −2.81 (−2.44) | −2.43 | 0.01 (0.06) | o1: ortho 1 | −2.94 | 0.03 |
| 2 | −2.14 (−1.77) | −1.75 | 0.15 (0.22) | o2: ortho 2 | −1.93 | 0.23 |
| 3 | −1.91 (−1.55) | −1.55 | 0.22 (0.29) | p1: para | −1.37 | 0.29 |
| | | | | m1: meta 1 | −1.02 | |
| | | | | m2: meta 2 | −1.02 | |

**Table 1. Binding energies and barriers for the first and second hydrogenation reactions.**

*First hydrogenation.* The first hydrogenation of the coronene cation can occur at three distinct locations as shown in black in Fig. 2, left panel: 1) an outer edge carbon atom (site 1 in black, 12 equivalent sites, $CH_2$ formation); 2) an inner edge carbon atom (site 2 in black, 6 equivalent sites, CH formation); 3) a center ring C atom (site 3 in black, 6 equivalent sites, CH formation). For the three hydrogenation sites considered here, the obtained zero point energy-corrected binding energies are reported in Table 1. The first hydrogenation reaction corresponds to a radical-radical reaction and leads to a closed shell system. As a consequence, large binding energies are obtained. Our results are in excellent agreement with previous calculations obtained with the hybrid B3LYP functional[21]. For the three cases, the zero point energy (ZPE) correction is about 0.4 eV (Table 1). Subsequent hydrogenations will correspond to one of the three prototypical former cases, and the ZPE correction for the binding energy will correspond to a global offset.

The hydrogenation barriers have been determined by a transition state calculation (see Table 1 for the ZPE corrected data). For H attachment on site 1, the transition state calculation gives a relatively low barrier of 0.01 eV and accompanies a binding energy, which is 0.7 eV higher than for the other two sites. This barrier is related to the out of plane motion of the existing CH bond in order to form the $CH_2$ group (see Fig. 2). Even though the formation of the $CH_2$ group weakens the $\pi$ system of the associated ring, the aromatic carbon backbone is only weakly perturbed by the H addition and remains planar. H attachment on site 3 implies a stronger perturbation of the carbon backbone with the active C carbon atom being puckered out of the aromatic plane. As a consequence, the barrier is much higher for this site (0.22 eV). H attachment on site 2 is an intermediate case with a barrier of 0.15 eV. The H atom attachment reactions are exothermic and the transition state geometries are close to the geometry of the reactants (early transition state). For the most stable final product, the barrier is the lowest.

Because of the large binding energy and the low barrier, the hydrogenation on site 1 is the most probable and the $C_{24}H_{13}^+$ cation (site 1) is expected to be rapidly formed and to be particularly abundant.

*Second hydrogenation.* The first hydrogenation leaves the coronene $\pi$ electron system weakened. The second hydrogenation has been investigated for all available sites of the same ring (see the different sites in red in Fig. 2, with the first H atom attached in position 1). The second hydrogenation reaction is a closed shell-radical reaction, resulting in a radical system. As a consequence, binding energies are typically smaller than for the first hydrogenation, with the exception of the ortho1 site (see Table 1), for which a slightly higher value is found. This result implies that additional electronic effects, such as hyperconjugation, contribute to the $C_{24}H_{14}^+$ cation stabilization. Even though, attachment to the ortho 1 site is energetically more favorable by 1 eV ($E_b = -2.94$ eV), the corresponding hydrogenation barrier amounts to 0.03 eV. This barrier on site o1, is much lower (by at least 0.2 eV) than for the o2 and p1 sites. The geometry, presented in Fig. 2 (right), shows the change of hybridization of the involved carbon and the torsion of the C-C bond to obtain a staggered ethane-like conformation. The aforementioned barrier of 0.03 eV is related to this local deformation and therefore is higher than for the first hydrogenation reaction. Because of the presence of this barrier, accumulation of $C_{24}H_{13}^+$ molecules is expected before further hydrogenations.

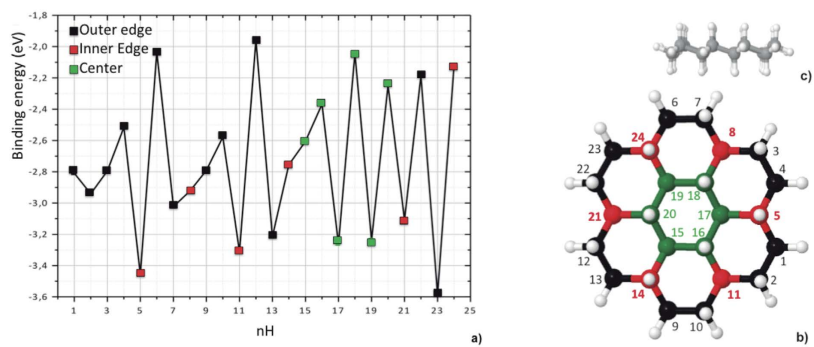

**Figure 3.** Panel (**a**) Binding energy of each H atom as a function of the number of H atoms adsorbed on the coronene cation. Panel (**b**) Fully hydrogenated coronene cation and the sequence which is followed for the addition of each hydrogen atom. The colors correspond to the location of the carbon, outer edge (black), inner edge (red) center ring(green). Panel (**c**) Geometry of the fully hydrogenated coronene cation.

*Further hydrogenation states.* Subsequent hydrogen attachment proceeds along a pathway involving the most stable cations. For the particularly stable species with $n_H = 5, 11, 17$ which accumulate in the experiment, hydrogenation binding energies are presented in Fig. 3a). The binding energies range between $-1.95$ and $-3.6$ eV and depend strongly on the number of additional hydrogens ($n_H$). The addition of hydrogen atoms, from the third up to the 24$^{th}$, follows the sequence shown in Fig. 3b. From the third up to the 14$^{th}$ hydrogenation, H attachment follows a cycle with 3 different locations 1) outer edge carbon atom on an adjacent aromatic ring, 2) adjacent outer edge carbon atom, 3) inner edge carbon atom belonging to two outer hydrogenated rings. After the 14$^{th}$ hydrogenation, five contiguous outer rings of the coronene cation have lost their aromaticity. From the 15$^{th}$ up to the 20$^{th}$ hydrogenations, the H atoms are bound to the inner ring, and for the last four hydrogenations, the H atoms are bound to the last outer ring.

If the H attachment involves two radicals, the product ($C_{24}H_{12+n_H}^+$, with odd $n_H$) has closed shell configuration with H binding energies below $-2.6$ eV. Furthermore, if the involved reaction does not imply a strong carbon backbone deformation, the reaction is expected to be barrier-less or associated with a low barrier (typically of the order of the first hydrogenation reaction i.e. 0.01 eV). This is the case for $n_H = 1, 3, 9$. If the reaction occurs between a radical and a closed shell system, the product ($C_{24}H_{12+n_H}^+$, with even $n_H$), is a radical, the binding energy is above $-3$ eV. Furthermore, if the involved reaction implies a substantial deformation of the carbon backbone, a larger barrier is expected (typically of the order of the second hydrogenation barrier of 0.03 eV). This is the case for $n_H = 4, 10$. Note that the binding energies are not always alternating from odd to even nH since they depend on the type of reactions (radical + radical or radical + closed shell system) but also on the deformation of the system.

Such alternating hydrogenation pattern leads to equilibrium structures for which the energy strains associated to the tetrahedrisation of the C atoms are minimized (C atoms change from trigonal planar sp$^2$ to tetrahedral sp$^3$ upon hydrogenation). One can notice that after hydrogenation the edge structure is alkane-like while the global structure is graphane-like, each $C_6$ ring having a chair conformation, as shown in Fig. 3c.

The largest differences in binding energies occur between the 5$^{th}$ to the 6$^{th}$, the 11$^{th}$ to the 12$^{th}$, the 17$^{th}$ to the 18$^{th}$ and the 23$^{th}$ to the 24$^{th}$ hydrogenations. The 5$^{th}$ hydrogenation ($C_{24}H_{16}^+$ + H, a radical + radical reaction) results in the formation of a closed-shell cation with 18 $\pi$ electrons distributed over 5 aromatic adjacent rings (this preserved aromatic part of the cation is a PAH-like $C_{19}H_{11}^{+[22]}$). This reaction is expected to have a small barrier (0.01 eV). The 6$^{th}$ hydrogenation ($C_{24}H_{17}^+$ + H, a radical + closed-shell reaction) leads to the formation of a radical. For this reaction, the transition state calculation gives a larger barrier of 0.11 eV. Consequently the $C_{24}H_{17}^+$ cation is expected to be prominent during successive hydrogenations. Identical mechanisms apply to the 11$^{th}$ and 12$^{th}$ (barrier height 0.17 eV) as well as 17$^{th}$ and 18$^{th}$ (barrier height 0.32 eV) hydrogenation reactions. For the 23$^{th}$ hydrogenation, the radical character of the cation $n_H = 22$ is concentrated on two C atoms and hydrogenation occurs at the 23$^{th}$ site (see Fig. 3b).

## Discussion

Experimentally, the hydrogenation of coronene cations shows the predominance of cations with 5, 11 as well as 17 extra hydrogens, which we refer to as magic numbers. This experimental observation is fully explained theoretically by the specific hydrogenation sequence derived from binding energies and attachment barriers. This sequence shows that the 6$^{th}$, 12$^{th}$, and 18$^{th}$ H atom attachment require substantially high barriers (larger than 0.1 eV). These barriers, guiding the H atoms into the most stable adsorption sites, explain the presence of the most probable cations that are observed experimentally as magic numbers.

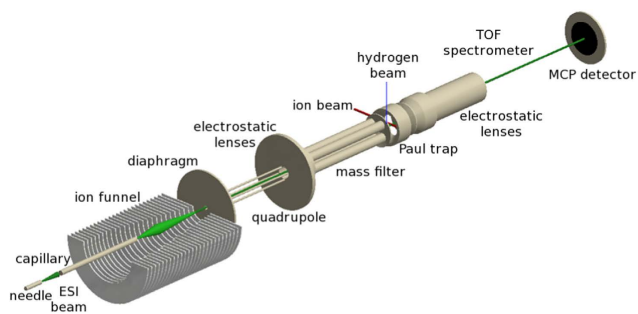

**Figure 4. The setup used, with the ion funnel, quadrupoles, ion trap, hydrogen source and detector.**

For neutral coronene, DFT calculations have shown that the first hydrogenation of neutral coronene is associated with a barrier (~60 meV), but that subsequent hydrogenation barriers vanish[23]. In this work the hydrogenation reactions have been studied up to $n_H = 8$. Our results indicate that the hydrogenation pattern is different already for $n_H = 3$ between the neutral and cation.

The sequence to hydrogenate coronene cations has implications for larger systems. The specific sequence derived in this study shows that coronene cations with a magic number of H atoms attached are more stable systems, and that fully hydrogenated coronene cations have $C_6$ rings with chair conformation (see Fig. 3c), which is similar to graphane[11]. Attaching atomic hydrogen to each site of the graphene lattice changes the hybridization from $sp^2$ into $sp^3$, thus removing electrons from the conducting $\pi$-bands and opening an energy gap, whose size depends on the H coverage[24]. To control the extent of graphene conductivity, selective hydrogen adsorption needs to be achieved. In this work, by using coronene as a model system for graphene related materials, we show that stable hydrogenation states allow to control the number of H atoms attached as well as their localization.

The discovery of magic numbers of H attached to PAHs has implications concerning the amount of aromatic and aliphatic carbon material that can be found in space. Observations of the aliphatic C-H band at 6.86 $\mu$m and the aromatic C-C band at 7.7 $\mu$m indicate that the fraction of carbon in PAHs which is in aliphatic form is $\leq 15\%$[25] (and with high percentage of cations in regions shielded from UV radiation[26,27]). Recent observations of the aromatic C-H band at 3.3 $\mu$m, as well as the aliphatic C-H band at 3.4 $\mu$m (which is associated either to methyl sidegroups attached to the PAHs, or to superhydrogenated PAHs[28]) show that the intensity of these bands change along the line-of-sight as a function of visual extinction[29]. These changes suggest that very small grains are photo-processed and lead to the production of PAHs with aliphatic sidegroups. Recent theoretical studies suggest that in environments subjected to strong UV radiation, small PAHs are likely to be mostly de-hydrogenated, while very large PAHs are likely to be super-hydrogenated[30]. Our study indicates that the sequence to hydrogenate PAHs as well as the presence of barriers that lead to magic numbers, will allow only for a certain degree of hydrogenation. This is consistent with the observations of 3.4 $\mu$m and 3.3 $\mu$m, showing that the absolute degree of implied excess hydrogenation is not large, and that the molecular population is still dominated by aromatic carbon, not aliphatic carbon[31].

Molecular hydrogen is the most abundant molecule in our Universe. Observations of many astrophysical environments revealed that this molecule can form and survive in extreme conditions such as strong UV/Xrays fields, cold and warm gas and dust, or post-shock regions. The formation of $H_2$ is therefore efficient in a wide range of astrophysical conditions, and in this, PAHs can be important interstellar catalysts. Molecular hydrogen, can form in interstellar space by involving hydrogen atoms attached to PAHs[1,33–35], and to PAH cations in dense regions ($\geq 10^3$ cm$^{-3}$) submitted to strong UV radiation ($\geq 10$ erg s$^{-1}$ cm$^{-2}$)[26,27]. Therefore, the formation of $H_2$ in space depends on the hydrogenation state and stability of the neutral and cationic PAHs. In this study, we show that hydrogenation of PAH cations lead to the occurrence of magic numbers and that collisions of PAHs with hydrogen atoms lead to hydrogenation, while de-hydrogenation by collision is not observed. Since the barriers for abstraction (reaction between an H atom from the gas phase and an H attached to a PAH to form $H_2$) are supposed to be small ($\leq 10$ meV for abstraction of H on a neutral coronene[23]), the sequence to hydrogenate PAHs leading to super-hydrogenated PAHs is crucial for the formation of $H_2$.

## Methods

**Experiments.** The hydrogenation of coronene cations in the gas phase has been studied experimentally[33,35]. The experiments have been performed using a home-built tandem-mass spectrometer shown schematically in Fig. 4 [35,36]. A beam of singly charged coronene radical cations ($[C_{24}H_{12}]^+$, m/z = 300) was extracted from an electrospray ion source. The ions were phase-space compressed in an RF ion funnel and subsequently in an RF quadrupole ion guide. Mass selection was accomplished by using an RF quadrupole mass filter. Accumulation of the ions took place in a three-dimensional RF ion trap (Paul trap). A He buffer gas at room temperature was used to collisionally cool the trapped cations. Exposure to gas-phase atomic hydrogen for variable periods of time led to multiple hydrogen adsorptions on the coronene cations. The density of atomic hydrogen in our experiments is of the order of $10^{11} - 10^{12}$ cm$^{-3}$. An electric extraction field was then applied between the trap end-caps to extract the trapped hydrogenated cations into a time-of-flight (TOF) mass spectrometer with resolution M/$\Delta$M ~ 300. To obtain mass spectra of sufficient statistics, typically hundreds of TOF traces were accumulated.

Electrospray ionization allows to gently transfer ions from the solution phase into the gas phase. The electrospray ion source was operated using a solution consisting of 600 μL of saturated solution of coronene in methanol, 350 μL of High-performance liquid chromatography (HPLC) grade methanol and 50 μL of 10 mM solution of $AgNO_3$ solution in methanol[37]. In the liquid phase, electron transfer from a coronene molecule to a silver ion leads to formation of the required radical cation.

The trapped ions are exposed to hydrogen atoms produced from $H_2$ by a Slevin-type hydrogen source which has been extensively used in crossed beam experiments[38,39]. While in the earlier work the dissociation fractions were determined by means of electron impact excitation or HeII line emission, we now use charge removal and dissociation induced by 40 keV $He^{2+}$. In this way we determine a hydrogen dissociation fraction of $n(H)/(n(H) + n(H_2)) \approx 0.3$. The temperature of the H beam is around room temperature (~25 meV).

After irradiating the coronene ions for a certain amount of time, an electric field is used to extract the trap content into the TOF mass spectrometer.

**DFT calculations.** Equilibrium geometries, binding energies and transition states have been obtained using the DFT method. Calculations have been performed with the ADF program, version 2013[40]. Spin polarized calculations have been performed for these open-shell systems with the modified PW1K hybrid functional[20] and at ζp (triple zeta polarized) basis set of Slater type orbitals. Numerical convergence for geometry is obtained with gradients < 0.001 Hartree per Å and distances are converged to 0.01 Å. Numerical convergence for energies is $10^{-6}$ hartree. Transition states are obtained as stationary points on the energy surface, and are characterized by a Hessian matrix with a negative eigenvalue associated with a motion along the reaction coordinate. For the super-hydrogenated cation $C_{24}H_{12+n}^+$, the H binding energy $\Delta E_b$ and barrier height $\Delta E^\#$ are calculated according:

$$\Delta E_b = E\left(C_{24}H_{12+n}^+\right) - E\left(C_{24}H_{12+(n-1)}^+\right) - E(H) \tag{1}$$

and

$$\Delta E^\# = E\left(C_{24}H_{12+n}^{+\#}\right) - E\left(C_{24}H_{12+(n-1)}^+\right) - E(H) \tag{2}$$

where $E(C_{24}H_{12+n}^+)$ and $E(C_{24}H_{12+n}^{+\#})$ are the equilibrium and transition state energies of the $C_{24}H_{12+n}^+$ cation. $E(C_{24}H_{12+(n-1)}^+)$ and $E(H)$ are the equilibrium energies of the previous $C_{24}H_{12+(n-1)}^+$ cation and atomic hydrogen respectively.

Previous calculations on the coronene system have been performed using the PW91 GGA functional for the neutral[23,41] and with the hybrid functional B3LYP for the cation[21]. For such systems, hybrid functional have been shown to better perform than the PW91 GGA functional which underestimate barrier heights and overestimate binding energies[42]. Moreover, MPW1K has been shown to predict accurate barrier heights and reaction energies[20]. The barriers have been obtained via a transition state calculation. The transition state is obtained has a saddle point along the reaction coordinate and is characterized by an imaginary frequency. The barriers obtained in the present work correspond to the attachment of an H atom coming from the gas phase.

### Acknowledgements

S.C. and L.B. are supported by the Netherlands Organization for Scientific Research (NWO; VIDI project 639.042.017). S.C. is supported by the European Research Concil (ERC; project PALs 320620). GR is supported by the Netherlands Organization for Scientific Research (NWO) (Dutch Astrochemistry Network & Rubicon 68-50-1410). The calculation of this work was performed using high performance computing resources provided by the Grant Equipement National de Calcul Intensif (GENCI)-Institut du Developpement et des Ressources en Informatique Scientifique (IDRIS) project 100331.

### Author Contributions

S.C. wrote the main manuscript. The experiments and analysis were performed by L.B., G.R., T.S., R.H., S.C. and M.S. The theoretical calculations were performed by N.R., S.M. and D.T.B. All authors reviewed the manuscript.

### Additional Information

**Competing financial interests:** The authors declare no competing financial interests.

**How to cite this article**: Cazaux, S. *et al.* The sequence to hydrogenate coronene cations: A journey guided by magic numbers. *Sci. Rep.* **6,** 19835; doi: 10.1038/srep19835 (2016).